\documentclass[nofootinbib,notitlepage,superscriptaddress,10pt,aps,prd,twocolumn]{revtex4-1}
\usepackage[utf8]{inputenc}
\usepackage{amsmath,mathtools}
\DeclareMathOperator{\sgn}{sign}
\usepackage{tensor}
\usepackage{amssymb}
\usepackage{graphicx}
\usepackage{lipsum}  
\usepackage{verbatim}
\newcommand{\leri}[1]{\left(#1\right)}

\newcommand{\abs}[1]{\left|#1\right|}

\begin{document}
\title{Scalar modes in extended hybrid metric-Palatini gravity: weak field phenomenology}
\author{Flavio Bombacigno}
\email{flavio.bombacigno@uniroma1.it}
\noaffiliation
\author{Fabio Moretti}
\email{fabio.moretti@uniroma1.it}
\affiliation{Physics Department, ``Sapienza'' University of Rome, P.le Aldo Moro 5, 00185 (Roma), Italy}
\author{Giovanni Montani}
\email{giovanni.montani@enea.it}
\affiliation{Physics Department, ``Sapienza'' University of Rome, P.le Aldo Moro 5, 00185 (Roma), Italy}
\affiliation{ENEA, Fusion and Nuclear Safety Department, C. R. Frascati,
	Via E. Fermi 45, 00044 Frascati (Roma), Italy}
\begin{abstract}
We investigate the nature of additional scalar degrees of freedom contained in extended hybrid metric-Palatini gravity, outlining the emergence of two coupled dynamical scalar modes. In particular, we discuss the weak field limit of the theory, both in the static case and from a gravitational waves perspective. In the first case, performing an analysis at the lowest order of the post parameterized Newtonian (PPN) structure of the model, we stress the settling of Yukawa corrections to the Newtonian potential. In this respect, we show that one scalar field can have long range interactions and used in the principle for mimicking dark matter effects. Concerning the  gravitational waves propagation, instead, we demonstrate that is possible to have well-defined physical degrees of freedom, provided by suitable constraints on model parameters. Moreover, the study of the geodesic deviation points out the presence of breathing and longitudinal polarizations due to these novel scalar waves, which on peculiar assumptions can give rise to beating phenomena during their propagation.
\end{abstract}

\maketitle
\section{Introduction}\label{sec1}
During last years, modified theories of gravity have been intensively studied in order to address problems of modern cosmology. Indeed, current evidences of a phase of accelerated expansion for the Universe \cite{Riess:1998cb,Perlmutter:1998np,Knop:2003iy,Amanullah:2010vv,Weinberg:2012es}, along with structure dynamics in astrophysical scenarios, e.g. galaxy rotation curves or
clusters properties \cite{Persic:1995ru,Wu:1998ju,Firmani:2000ce}, represent inescapable issues for any reliable attempt of providing a unitary theoretical picture of gravitational interaction on different scales.
In fact, our present description of Universe evolution, based on the so-called $\Lambda$CDM model, requires the uncomfortable introduction of two unspecified dark components into the matter-energy budget of the Universe. Dark energy, responsible for a de Sitter phase of accelerated expansion and comparable with a cosmological constant term in Einstein equations ($\Lambda$), and cold dark matter (CDM), thought as non relativistic particles interacting with ordinary matter mostly gravitationally \cite{Overduin:2004sz,Baer:2014eja,Bernal:2017kxu}. This model however, even though phenomenologically well-grounded, is not capable of a satisfactory theoretical justification for its additional dark elements. Especially, it is still object of debate the process originating the effective cosmological constant \cite{Martel:1997vi,Carroll:2000fy,Peebles:2002gy,Padmanabhan:2002ji}, whose observed value is in contrast with predictions of quantum field theory, or the proper nature of dark matter particles \cite{Navarro:1995iw,Jungman:1995df,Bertone:2004pz,ArkaniHamed:2008qn}. With this regard, a different perspective is therefore offered by the possibility of modifying the nature of the gravitational interaction as predicted by General Relativity, with the aim of accounting for these exotic phenomena as purely dynamical effects, e.g. introducing additional degrees of freedom \cite{Nojiri:2005jg,DeLaurentis:2015fea,Joyce:2016vqv} or modified stress-energy couplings to geometry \cite{Nicolis:2008in,Deffayet:2009wt,Harko:2011kv,Odintsov:2013iba,Wu:2018idg,Barrientos:2018cnx}. Of course, a large number of choices for an extended Einstein-Hilbert action is actually feasible, involving different contributions in metric derivatives \cite{Bergmann:1968ve,Lovelock:1971yv,Horndeski:1974wa,Whitt:1984pd,Schmidt:2006jt,Bahamonde:2015zma}, as well as gauge theory approaches for the gravitational field \cite{Hehl:1994ue,Jackiw:2003pm,Blagojevic:2012bc,Hehl:2013qga,Ashtekar:1986yd,Immirzi:1996di}. Among these available models, $f(R)$ theories stand for their relevance and simplicity \cite{Sotiriou:2008rp}, where a new degree of freedom is introduced by replacing the Ricci scalar $R$ of the standard General Relativity action with a generic function of it, leading to fourth order equations of motion for the metric field. Cosmological scenarios stemming from such revisited theoretical framework have been deeply investigated, with dark energy-like solutions widely discussed \cite{Capozziello:2005ku,Cognola:2007zu,Nojiri:2010wj,Nojiri:2017ncd}, and dark matter issue addressed by means of the additional scalar mode featuring this reformulation \cite{Boehmer:2007kx,Stabile:2013jon}, made manifest in its scalar tensor restatement \cite{Flanagan:2003iw,Olmo:2005hc,Capone:2009xk,ST}. In this respect, however, the requirement of preserving solar system local dynamics \cite{Will:2005va,Zakharov:2006uq,Chiba:2006jp,Schmidt:2008qi,Berry:2011pb}, consisting in very short range scalar interaction, turned out to be inconsistent with demands of late time expansion, involving instead astrophysical range deformations of gravitational force, and led to introduce peculiar screening mechanisms \cite{Khoury:2003rn,Brax:2008hh,Capozziello:2007eu}.
\\ \indent Another source of ambiguity for the gravitational field dynamics is offered by the nature of the metric field and the affine connection, which could be considered in principle as independent variables. Such an approach, corresponding to the so-called Palatini (or first order) formulation, appears very promising especially for its implication in the quantization of gravity as a gauge field theory \cite{Ashtekar:1986yd,Immirzi:1996di}. However, even if the Hilbert-Palatini action is in vacuum at all equivalent to the metric Einsten-Hilbert analogous \cite{Holst:1995pc}, it outlines significant differences for instance when fermions are included in the dynamics \cite{Mercuri:2007ki}. In fact, spinor fields couple to connection and induce a non-vanishing torsion in spacetime structure, so that the equivalence with the second order approach is intrinsically lost, forcing us to deal instead with Einstein-Cartan geometry \cite{Hehl:1976kj,Shapiro:2001rz}. Similar issues arise when Palatini scheme is applied to $f(\mathcal{R})$ models \cite{Olmo:2011uz}, and several discrepancies emerge with respect to the corresponding metric (or second order) analysis. Especially, the connection turns out to be an auxiliary field devoid of proper dynamics, whose expression depends on the form of the function $f(\cdot)$, and Palatini $f(\mathcal{R})$ gravity can be conveniently restated as a metric theory endowed with torsion \cite{Olmo:2011uz,Bombacigno:2018tbo,Bombacigno:2018tih}. Therefore, the additional scalar degree is not dynamical, but it affects the way matter sources and spacetime curvature interact, and also in vacuum the two reformulations are not equivalent, being Palatini case featured by an effective cosmological constant, inherently related to the form of the $f(\cdot)$ function. 
\\ \indent As originally proposed in \cite{Harko:2011nh}, an intriguing perspective is constituted by the possibility of combining both the approaches, considering actions which contain Palatini $f(\mathcal{R})$ modifications to ordinary Einstein-Hilbert metric Lagrangian. Particularly, these theories successfully accomplish the result of providing long range scalar mode, able to reproduce dark matter effects \cite{Capozziello:2013uya,Capozziello:2012qt,Capozziello:2013yha}, without violating solar system observational constraints and invoking the so-called \textquotedblleft chameleon mechanism\textquotedblright \cite{Khoury:2003rn, Brax:2008hh,Capozziello:2007eu}. Furthermore, cosmological solutions have been extensively investigated, obtaining accelerated expansion scenarios \cite{Carloni:2015bua,Leanizbarrutia:2017xyd}, and studies about compact objects and spherically symmetric static configuration have been performed \cite{Danila:2016lqx,Danila:2018xya}.
\\ Here, we deal with a further generalization of these mixed models, and consider a scalar action as in \cite{Tamanini:2013ltp,Rosa:2017jld}, where the function $f$ is assumed to depend on both the Ricci scalars, metric and Palatini ones. Especially, we analyze in detail the features of the theory in the weak field limit. It is easy to recognize that in such a type of theory, the scalar-tensor representation is still possible, but now two distinct scalar degrees of freedom come out. These non-minimally coupled scalar fields are dynamically characterized by the form of the potential term they obey as a result of the form of the original function $f({R,\mathcal{R}})$. Critical points of the potential (minima, maxima and saddle points) are relevant for the
local gravitational field dynamics, as it is concerned in the PPN limit or when the propagation of gravitational waves is taken into account.
\\ We analyze situation in which the two emerging massive scalar modes have, in the diagonal representation, well-defined masses, ruling out of the theory the non-physical situations in which tachyon modes are present (see \cite{Koivisto:2013kwa} for a comparison).
\\ On the level of PPN analysis, we show how General Relativity can be still recovered with high degree of precision in the Solar System, as far as the theory parameters are suitably constrained, also in the presence of long range scalar interaction, which can be adopted in principle to reproduce dark matter effects. Then, we analyze the propagation of the gravitational waves in the presence of the two additional massive scalar modes. The deformation of the standard wave polarizations is investigated in some detail for a rather general spanning of the parameter space. In particular, we discuss the intriguing case of nearly degenerate massive modes, and we study the very peculiar phenomenon of wave beating. Such a beating mode is a very striking track of the considered modified theory of gravity and it suggests that upper limits on the existence of mixed $f(R,\mathcal{R})$ model can be experimentally put via present and incoming interferometer devices \cite{TheLIGOScientific:2016src,Abbott:2017tlp,Abbott:2018utx,Capozziello:2008rq,Chatziioannou:2012rf,Isi:2015cva,Maselli:2016ekw,Zhang:2017sym,Blaut:2019fxb}.
\\ The paper is organized as follows. In Sec.~\ref{sec2} extended hybrid metric-Palatini models are briefly discussed and their scalar tensor representation introduced; in Sec.~\ref{sec3} we analyze the first PPN corrections in the static weak field limit, pointing out the appearing of Yukawa corrections to gravitational potential given by both the additional scalar degrees; in Sec.~\ref{sec4} we address the propagation of gravitational waves in vacuum, investigating to some extent the theory structure in order to have well-defined physical modes; in Sec.~\ref{sec5} we study the effects on geodesic deviation equation of scalar fields, tracing analogies with metric $f(R)$ theories, and discussing the settling of beating phenomena; in Sec.~\ref{sec6} we refine the analysis of Secs.~\ref{sec3} and \ref{sec4} putting several constraints on the form of the function $f(R,\mathcal{R})$; in Sec.~\ref{sec7} conclusions are drawn.

\section{Extended hybrid metric-Palatini theories}\label{sec2}
Formerly introduced in \cite{Flanagan:2003iw,Tamanini:2013ltp}, extended hybrid metric-Palatini theories are described by the action\footnote{We set $\kappa=8\pi G$ and $c=1$.}
\begin{equation}
S=\frac{1}{2\kappa}\int d^4x\;\sqrt{-g}\,f(R,\mathcal{R}) +S_M
\label{f(R)action1}
\end{equation}
where $S_M$ stands for the generic matter contribution and the function $f(R,\mathcal{R})$ is assumed to depend on both the metric and Palatini Ricci scalars, denoted by $R$ and $\mathcal{R}$, respectively. Accordingly, we deal with two different kind of affine connections, i.e. the standard Levi-Civita connection related to the metric curvature scalar\footnote{We adopted the mostly plus spacetime signature $(-1,+1,+1,+1)$ and the following convention for the Riemann tensor: $R\indices{^\mu_{\nu\rho\sigma}}=\partial_{\rho}\Gamma\indices{^\mu_{\nu\sigma}}-\partial_{\sigma}\Gamma\indices{^\mu_{\nu\rho}}+\Gamma\indices{^\mu_{\tau\rho}}\Gamma\indices{^\tau_{\nu\sigma}}-\Gamma\indices{^\mu_{\tau\sigma}}\Gamma\indices{^\tau_{\nu\rho}}$, with $R_{\mu\nu}=R\indices{^\rho_{\mu\rho\nu}}$.} $R$:
\begin{equation}
    R=g^{\mu\nu}R_{\mu\nu}(\Gamma(g)),
\end{equation}
with
\begin{equation}
    \Gamma\indices{^\rho_{\mu\nu}}=\frac{g^{\rho\tau}}{2}\leri{\partial_\mu g_{\nu\tau}+\partial_\nu g_{\mu\tau}-\partial_\tau g_{\mu\nu}},
\end{equation}
and the independent connection $\Tilde{\Gamma}\indices{^\rho_{\mu\nu}}$ defining the Palatini Ricci scalar
\begin{equation}
    \mathcal{R}=g^{\mu\nu}\mathcal{R}_{\mu\nu}(\tilde{\Gamma}).
\end{equation}
The form of $\Tilde{\Gamma}\indices{^\rho_{\mu\nu}}$ can be dynamically determined by evaluating its equation of motion from \eqref{f(R)action1}. Indeed, under the assumption that the matter fields only minimally couple with the metric, it results in
\begin{equation}
    \tilde{\nabla}_\rho\leri{\sqrt{-g}f_\mathcal{R}g^{\mu\nu}}=0,
    \label{defLeviconformal}
\end{equation}
where $f_\mathcal{R}\equiv \partial_{\mathcal{R}}f$ (similarly for $R$ and higher order derivatives), and $\tilde{\nabla}_\rho$ denotes the covariant derivative from $\tilde{\Gamma}\indices{^\rho_{\mu\nu}}$. \\ Especially, if we neglect the issue concerning the role of the torsion in Palatini $f(\mathcal{R})$ models (see \cite{Olmo:2011uz} for a review and \cite{Bombacigno:2018tbo,Bombacigno:2018tih} for specific applications), the solution of \eqref{defLeviconformal} is given by
\begin{equation}
    \tilde{\Gamma}\indices{^\rho_{\mu\nu}}=\frac{\tilde{g}^{\rho\tau}}{2}\leri{\partial_\mu \tilde{g}_{\nu\tau}+\partial_\nu \tilde{g}_{\mu\tau}-\partial_\tau \tilde{g}_{\mu\nu}},
    \label{levicivitaconform}
\end{equation}
which represents the Levi-Civita analougous for the conformal metric $\tilde{g}_{\mu\nu}\equiv f_\mathcal{R}\,g_{\mu\nu}$, and allows us to recast the Palatini Ricci scalar as
\begin{equation}
    \mathcal{R}=R+\frac{3}{2 f_\mathcal{R}^2}\partial_\mu f_\mathcal{R}\partial^\mu f_\mathcal{R}-\frac{3\Box f_\mathcal{R}}{f_\mathcal{R}}.
    \label{sviluppo R palatini f}
\end{equation}
Of course, since the function $f_{\mathcal{R}}$ can contain in principle both the curvature scalars, in \eqref{sviluppo R palatini f} is actually embedded a differential equation relating $R$ and $\mathcal{R}$. Therefore, in order to make definition \eqref{levicivitaconform} well-grounded, we have to provide a further relation between the metric Ricci scalar $R$ and the Palatini curvature $\mathcal{R}$. That can be accomplished by tracing the equation of motion for the metric field $g_{\mu\nu}$, i.e.
\begin{equation}
f_R\,R_{\mu\nu}+f_\mathcal{R}\,\mathcal{R}_{\mu\nu}-\frac{1}{2}g_{\mu\nu}f-(\nabla_\mu\nabla_\nu-g_{\mu\nu}\Box)f_R=\kappa T_{\mu\nu},
\label{eqmetrica}
\end{equation}
yielding to
\begin{equation}
    f_R\, R+f_\mathcal{R}\,\mathcal{R}-2f+3\Box f_R=\kappa T.
    \label{eqstrutturale}
\end{equation}
Then, since $f_R$ is function of both the metric and the Palatini scalars as well, relation \eqref{eqstrutturale} constitutes a second differential equation for $R$ and $\mathcal{R}$, which along with \eqref{sviluppo R palatini f} forms a set of highly coupled differential equations for the two different curvatures, here rewritten for the sake of clarity:
\begin{equation}
    \begin{cases}
    & 3\Box f_\mathcal{R}-f_\mathcal{R}R+f_\mathcal{R}\mathcal{R}-\frac{3}{2 f_\mathcal{R}}\partial_\mu f_\mathcal{R}\partial^\mu f_\mathcal{R}=0 \\
    & 3\Box f_R+f_R\, R+f_\mathcal{R}\,\mathcal{R}-2f=\kappa T.
    \end{cases}
    \label{system diff equation curvatures}
\end{equation}
This suggests that in extended hybrid models we actually deal with two independent additional degrees of freedom, somehow connected to the two type of scalar curvatures the theory is equipped with. It is worth noting that this property depends crucially on the form of \eqref{eqstrutturale}, with special focus on $f_R$ contributes. Indeed, in hybrid models originally discussed in \cite{Harko:2011nh} we simply have $f_R=1$, and \eqref{eqstrutturale} reduces to an algebraic constraint relating the Palatini curvature both to the metric Ricci scalar and the trace of the stress energy tensor, i.e.
\begin{equation}
    f_\mathcal{R}\,\mathcal{R}-2f=\kappa T+R,
    \label{structural equation hybrid}
\end{equation}
which can be solved in principle for $\mathcal{R}=\mathcal{R}(R,T)$. This in turn implies that the first of \eqref{system diff equation curvatures} boils down to a differential equation for the metric scalar R, in the presence of non trivial stress energy source terms, and we just retain an additional degrees of freedom. We point out that such an outcome is to some extent preserved also in extended hybrid theories, when metric and Palatini terms in the function $f$ are actually separable, i.e. $f(R,\mathcal{R})=f_1(R)+f_2(\mathcal{R})$. In this case, in fact, constraint \eqref{structural equation hybrid} generalizes to
\begin{equation}
    f_{2\mathcal{R}}\,\mathcal{R}-2f_2=\kappa T+2f_1-3\Box f_{1R}-f_{1R}\,R,
\end{equation}
and the Palatini scalar $\mathcal{R}$ turns out to depend also on the derivatives of $R$, i.e. $\mathcal{R}=\mathcal{R}(R,\nabla R, \Box R,T)$. Again, the only additional dynamical degree is still $R$, even though its evolution is now encoded by a higher order differential equation.
\\In the following we will restrict our attention to the general case $f_{R\mathcal{R}}\neq 0$, where $R$ and $\mathcal{R}$ represent truly independent degrees of freedom, whose dynamics is described by fourth-order differential equations as displayed in \eqref{system diff equation curvatures}.

\subsection{Scalar-tensor formulation}
As discussed in \cite{Tamanini:2013ltp}, if the determinant of the Hessian matrix for $f(R,\mathcal{R})$ is not vanishing, i.e.
\begin{equation}
    f_{RR}f_{\mathcal{RR}}-f_{R\mathcal{R}}^2\neq 0,
\end{equation}
action \eqref{f(R)action1} can be rearranged in the scalar-tensor form
\begin{equation}
    S=\frac{1}{2\kappa}\int d^4x\;\sqrt{-g}\,\leri{\psi R+\xi \mathcal{R}-V(\psi,\xi)}+S_M,
    \label{scalartensoraction1}
\end{equation}
where we introduced the scalar fields $\psi\equiv f_R,\;\xi\equiv f_\mathcal{R}$, together with the potential term
\begin{equation}
    V(\psi,\xi)\equiv \psi R(\psi,\xi)+\xi \mathcal{R}(\psi,\xi)-f(R(\psi,\xi),\mathcal{R}(\psi,\xi)).
    \label{potential V}
\end{equation}
Hence, taking into account \eqref{sviluppo R palatini f} and the definition of $\xi$,
\eqref{scalartensoraction1} can be rearranged in its gravitational part as
\begin{equation}
    S_g=\frac{1}{2\kappa}\int d^4x\;\sqrt{-g}\,\leri{(\psi+\xi)R+\frac{3}{2\xi}\partial_\mu\xi\partial^\mu\xi-V(\psi,\xi)},
    \label{scalartensoraction2}
\end{equation}
Then, defining a new scalar field $\phi=\psi+\xi$, we can finally rewrite \eqref{scalartensoraction2} in the form
\begin{equation}
     S_g=\frac{1}{2\kappa}\int d^4x\;\sqrt{-g}\,\leri{\phi R+\frac{3}{2\xi}\partial_\mu\xi\partial^\mu\xi-W(\phi,\xi)},
\end{equation}
where $W(\phi,\xi)\equiv V(\phi-\xi,\xi)$ and only the scalar $\phi$ is coupled to the metric Ricci scalar.
\\ Then, varying \eqref{scalartensoraction2} with respect to the metric field we get
\begin{equation}
    \begin{split}
        R_{\mu\nu}-\frac{1}{2}g_{\mu\nu}R&-\frac{1}{\phi}(\nabla_\mu\nabla_\nu-g_{\mu\nu}\Box)\phi+\frac{3}{2\xi\phi}\partial_\mu\xi\partial_\nu\xi\\&-\frac{1}{2\phi}g_{\mu\nu}\leri{\frac{3}{2\xi}\partial_\rho\xi\partial^\rho\xi-W(\phi,\xi)}=\frac{\kappa}{\phi} T_{\mu\nu},
    \end{split}
    \label{equationmetricscalartensor}
\end{equation}
while equations for $\phi$ and $\xi$ are given by, respectively
\begin{align}
    & R=\frac{\partial W(\phi,\xi)}{\partial \phi}
    \label{equationphi}\\
    & \frac{3}{2\xi^2}\partial_\mu\xi\partial^\mu\xi-\frac{3\Box\xi}{\xi}-\frac{\partial W(\phi,\xi)}{\partial \xi}=0.
\label{equationxi}
\end{align}
Now, we can evaluate the trace of \eqref{equationmetricscalartensor}, resulting in
\begin{equation}
    R=\frac{3\Box \phi}{\phi}-\frac{3\partial_\mu\xi\partial^\mu\xi}{2\xi\phi}+\frac{2 W}{\phi}-\frac{\kappa T}{\phi},
    \label{traccia R}
\end{equation}
which plugged into \eqref{equationphi}-\eqref{equationxi} leads to the following set of coupled differential equations
\begin{align}
          & \Box \phi -\frac{1}{2\xi}\partial_\mu\xi\partial^\mu\xi+\frac{2W-\phi W_\phi}{3}=\frac{\kappa}{3} T \label{equationphi2}\\
          & \Box \xi -\frac{1}{2\xi}\partial_\mu\xi\partial^\mu\xi+\frac{\xi W_\xi}{3}=0 \label{equationxi2},
  \end{align}
where $W_\phi$ and $W_\xi$ are defined by analogy with $f_{R,\mathcal{R}}$. This set of equations is the scalar-tensor equivalent of \eqref{system diff equation curvatures}, with the additional scalar degrees of freedom now embodied in the independent fields $\phi,\,\xi$. With this respect, even if the transformation $\phi=\psi+\xi$ seems to artificially relate the degrees $\phi$ and $\xi$, it actually preserves the dynamical content of the theory, a fact that can be further appreciated by evaluating the equations of motion for the original fields $\psi$ and $\xi$ directly from \eqref{scalartensoraction1}, i.e.
\begin{align}
          & \Box \psi +\frac{2V-\psi V_\psi-\xi V_\xi}{3}=\frac{\kappa}{3} T \label{equationpsi3}\\
          & \Box \xi -\frac{1}{2\xi}\partial_\mu\xi\partial^\mu\xi+\frac{\xi( V_\xi-V_\psi)}{3}=0 \label{equationxi3}.
  \end{align}
This in turn guarantees that when interested in perturbation theory, the departure of $\phi$ and $\xi$ from background values could be considered truly independent, as long as they represent proper dynamical variables, as previously discussed when we look at \eqref{system diff equation curvatures}.
\section{Post parameterized Newtonian corrections}\label{sec3}
It is a well-established result (see \cite{Olmo:2005hc,Harko:2011nh}) that additional scalar degrees can remarkably affect the dynamics of gravitating system in weak field and slow motion case. In particular, Yukawa corrections are usually obtained for the Newtonian potential, and the requirement of reproducing local experiment results allows in general to put several constraints on theory parameters \cite{Zakharov:2006uq,Chiba:2006jp,Berry:2011pb}. In this regard, the easiest way of determining the effects of the fields $\phi$ and $\xi$ in a slightly curved spacetime is to consider a quasi-Minkowskian system of local coordinates where the metric can be put into the form
\begin{equation}
    g_{\mu\nu}\approx\eta_{\mu\nu}+h_{\mu\nu},
    \label{metric expans}
\end{equation}
with $\abs{h_{\mu\nu}}\ll 1$, and the scalar fields $\phi,\,\xi$ are given by
\begin{equation}
    \phi=\phi_0+\delta\phi\,\quad \xi=\xi_0+\delta\xi.
    \label{expansphixi}
\end{equation}
Here $\phi_0$ and $\xi_0$ represent background values fixed by cosmological boundary conditions, which evolve adiabatically in time according the cosmological background curvature. Local fluctuations are denoted by $\delta\phi,\,\delta\xi\sim \mathcal{O}(h)$, which we assume to vanish outside the region described by \eqref{metric expans}. Now, by virtue of \eqref{expansphixi} we can expand the potential $W(\phi,\xi)$ as:
\begin{equation}
\begin{split}
    W(\phi,\xi)\simeq  \,&W_0+W_{0,\phi}\,\delta\phi+W_{0,\xi}\,\delta\xi+\\
    &+\frac{1}{2}\leri{W_{0,\phi\phi}\delta\phi^2+W_{0,\xi\xi}\delta\xi^2+2W_{0,\phi\xi}\delta\phi\delta\xi},
\end{split}
\label{taylor expans W}
    \end{equation}
where the subscript $0$ denotes evaluation at the point $(\phi_0,\,\xi_0)$, which we require to be located in the neighbourhood of a stable minimum for $W(\phi,\xi)$, so that the smallness of corrections $\delta\phi$ and $\delta\xi$ be preserved by the dynamics. We assume therefore that
\begin{equation}
    \begin{split}
        &\det \text{H} (W_0)\equiv W_{0,\phi\phi}W_{0,\xi\xi}-W_{0,\phi\xi}^2 > 0,\\
        & W_{0,\phi\phi},\,W_{0,\xi\xi}>0.
        \label{ppn condition on W}
    \end{split}
\end{equation}
where we introduced $\text{H} (W_0)$, the Hessian matrix for the potential $W(\phi,\xi)$ evaluated at the point $(\phi_0,\xi_0)$. Moreover, given that the value $W_0$ is related to the background curvature by
\begin{equation}
    R^{(0)}=\frac{2W_0}{\phi_0},
\end{equation}
as it can be inferred by taking the lowest order in \eqref{traccia R}, quasi-Minkowskian conditions imply that locally we can set $R^{(0)}=\epsilon$, with $\epsilon$ a small parameter quantifying departure from spacetime flatness. In particular, since it turns out to be actually responsible for a divergent term in the PPN corrections to the gravitational potential (see later), we constraint it to be small enough so that its contribution is negligible on the considered scales, i.e. wherever \eqref{metric expans} is valid.
\\ Now, let us recast \eqref{equationmetricscalartensor} into the form
\begin{equation}
    \begin{split}
        R_{\mu\nu}=&\frac{\kappa}{\phi}\leri{T_{\mu\nu}-\frac{1}{2}g_{\mu\nu}T}+\frac{1}{2\phi}g_{\mu\nu}\leri{\Box \phi +W}\\
        &+\frac{1}{\phi}\nabla_\mu\nabla_\nu\phi-\frac{3}{2\phi\xi}\partial_\mu\xi\partial_\nu\xi,
    \end{split}
\end{equation}
which, once we fixed the Nutku gauge \cite{Nutku}
\begin{equation}
    h\indices{^\mu_{\nu,\mu}}-\frac{1}{2}h\indices{^\mu_{\mu,\nu}}=\frac{\partial_\nu\delta\phi}{\phi_0},
\end{equation}
gives at the lowest order in perturbation the following equations for the metric $h_{\mu\nu}$ components:
\begin{align}
    &\bigtriangleup\leri{h_{00}^{(2)}-\frac{\delta\phi}{\phi_0}}=-\frac{\kappa\rho}{\phi_0}+\frac{\epsilon}{2}
    \label{PPN h00}\\
    &\bigtriangleup\leri{h_{ij}^{(2)}+\delta_{ij}\frac{\delta\phi}{\phi_0}}=-\delta_{ij}\leri{\frac{\kappa\rho}{\phi_0}+\frac{\epsilon}{2}}
    \label{PPN hij},
\end{align}
where we neglected time derivatives, we set $\bigtriangleup=\nabla^2$ for the Laplacian operator and $T_{00}=-T\approx\rho$, $T_{ij}\approx 0$.
Analogously, we can rearrange \eqref{equationphi2} and \eqref{equationxi2} as
\begin{align}
    &(\bigtriangleup-m_\phi^2)\delta\phi+\frac{2W_{0,\xi}-\phi_0 W_{0,\phi\xi}}{3}\delta\xi=-\frac{\kappa}{3}\rho
    \label{PPNphi}\\
    &(\bigtriangleup-m_\xi^2)\delta\xi+\frac{\xi_0 W_{0,\phi\xi}}{3}\delta\phi=0
    \label{PPNxi},
\end{align}
with
\begin{equation}
    m_\phi^2\equiv\frac{\phi_0 W_{0,\phi\phi}-W_{0,\phi}}{3}\qquad m_\xi^2\equiv -\frac{\xi_0 W_{0,\xi\xi}+W_{0,\xi}}{3},
\end{equation}
and zero order terms satisfying
\begin{equation}
    W_{0,\phi}=\epsilon\qquad \xi_0 W_{0,\xi}=0.
    \label{PPNzaaro}
\end{equation}
Now, in order to solve \eqref{PPNphi}-\eqref{PPNxi} is useful to find a suitable change of variables with the aim of decoupling the equations of motion for $\delta\phi,\,\delta\xi$. That can be accomplished by introducing the matrix
\begin{equation}
    \text{A}\equiv
    \begin{pmatrix}
    m_\phi^2 & \frac{\phi_0\,W_{0,\phi\xi}-2W_{0,\xi}}{3} \\ -\frac{\xi_0\, W_{0,\phi\xi}}{3} & m_\xi^2
    \end{pmatrix},
    \label{def A}
\end{equation}
and the vectors
\begin{equation}
    \mathbf{\Phi}\equiv 
        \begin{pmatrix}
        \delta\phi  \\
        \delta \xi\end{pmatrix}\qquad \mathbf{T}\equiv -\frac{\kappa}{3}
        \begin{pmatrix}
        \rho  \\
        0
        \end{pmatrix},
\end{equation}
which allow us to rewrite the set \eqref{PPNphi}-\eqref{PPNxi} as
\begin{equation}
    (\text{I}_{2\times 2}\bigtriangleup-\text{A})\mathbf{\Phi}=\mathbf{T}
    \label{PPN matrix coupled system}
\end{equation}
where $\text{I}_{2\times 2}$ denotes the identity matrix of dimension 2.
\\Then, the system \eqref{PPN matrix coupled system} can be decoupled, that is A turned into diagonal form, by simply evaluating the matrix P of its eigenvectors. Therefore, let us rearrange \eqref{PPN matrix coupled system} like
\begin{equation}
(\text{I}_{2\times 2}\bigtriangleup-\text{A}_\text{D})\mathbf{\Phi_D}=\text{P}^{-1}\mathbf{T}
\label{PPN decoupled scalars}
\end{equation}
with $\text{A}_\text{D}\equiv \text{P}^{-1}\text{A}\text{P}$ diagonal and $\mathbf{\Phi_D}\equiv \text{P}^{-1}\mathbf{\Phi}$. We observe that the stress energy contributions to \eqref{PPNphi}-\eqref{PPNxi} are shuffled, so that we expect that matter sources could enter now both the equations for the decoupled scalar fields.
\\ Let us denote the elements of P and $\text{P}^{-1}$ with
\begin{equation}
    \text{P}=
    \begin{pmatrix}
    p_{11} & p_{12} \\
    p_{21} & p_{22}
    \end{pmatrix}
    \qquad
    \text{P}^{-1}=
    \begin{pmatrix}
    \bar{p}_{11} & \bar{p}_{12} \\
    \bar{p}_{21} & \bar{p}_{22}
    \end{pmatrix}
\end{equation}
and $\mathbf{\Phi}_D=(\delta\phi_D,\delta\xi_D)$. It is then possible to rewrite \eqref{PPN decoupled scalars} as
\begin{align}
    &(\bigtriangleup-M_\phi^2)\delta\phi_D=-\bar{p}_{11}\frac{\kappa}{3}\rho
    \label{PPN decoupled phi}\\
    &(\bigtriangleup-M_\xi^2)\delta\xi_D=-\bar{p}_{21}\frac{\kappa}{3}\rho
    \label{PPN decoupled xi},
\end{align}
where $M_{\phi,\xi}$ are the masses of the decoupled scalar fields which we require to be positive and that still have to be determined explicitly. Solutions for \eqref{PPN decoupled phi}-\eqref{PPN decoupled xi} can be easily obtained, i.e.
\begin{align}
    &\delta\phi_D(x)=\frac{2\bar{p}_{11}G}{3}\int d^3x'\,\frac{\rho(x')}{\abs{x-x'}}e^{-M_\phi\abs{x-x'}}
    \label{PPN phi solution}\\
    &\delta\xi_D(x)=\frac{2\bar{p}_{21}G}{3}\int d^3x'\,\frac{\rho(x')}{\abs{x-x'}}e^{-M_\xi\abs{x-x'}},
    \label{PPN xi solution}\\
\end{align}
where the integration is performed over the matter source. Hence, the solution for \eqref{PPN h00}-\eqref{PPN hij} can be written down, noting that the field $\delta\phi$ is actually a linear combination by means of P of the decoupled modes $\delta\phi_D,\,\delta\xi_D$, i.e.
\begin{equation}
    \delta\phi(x)=p_{11}\delta\phi_D(x)+p_{12}\delta\xi_D(x),
\end{equation}
which leads to
\begin{widetext}
\begin{align}
    &h_{00}^{(2)}(x)=\frac{2G}{\phi_0}\int d^3x'\,\frac{\rho(x')}{\abs{x-x'}}\leri{1+\frac{p_{11}\bar{p}_{11}e^{-M_\phi\abs{x-x'}}+p_{12}\bar{p}_{21}e^{-M_\xi\abs{x-x'}}}{3}}+\frac{\epsilon}{12}\abs{x-x_S}^2
    \label{PPN h00 solution}\\
    &h_{ij}^{(2)}(x)=\delta_{ij}\leri{\frac{2G}{\phi_0}\int d^3x'\,\frac{\rho(x')}{\abs{x-x'}}\leri{1-\frac{p_{11}\bar{p}_{11}e^{-M_\phi\abs{x-x'}}+p_{12}\bar{p}_{21}e^{-M_\xi\abs{x-x'}}}{3}}-\frac{\epsilon}{12}\abs{x-x_S}^2},
    \label{PPN hij solution}
\end{align}
\end{widetext}
where $x_S$ is an integration constant related to the source. Now, in spherically symmetric case and far away from the source, $\delta\phi(x)$ and the metric perturbations take the simpler form
     \begin{align}
        &\delta\phi(r)\simeq \frac{2G M_{\odot}}{3r}\leri{p_{11}\bar{p}_{11}e^{-M_\phi r}+p_{12}\bar{p}_{21}e^{-M_\xi r}}\\
        &h_{00}^{(2)}(r)\simeq \frac{2G_{eff}M_{\odot}}{r}+\frac{\epsilon}{12}r^2
        \label{symm far away h00}\\
        &h_{ij}^{(2)}(r)\simeq \delta_{ij}\leri{\frac{2\gamma G_{eff}M_{\odot}}{r}-\frac{\epsilon}{12}r^2},
        \label{symm far away hij} 
     \end{align}
with $M_{\odot}$ the Newtonian mass of the central body and the modified gravitational constant defined as
\begin{equation}
    G_{eff}\equiv\frac{G}{\phi_0}\leri{1+\frac{p_{11}\bar{p}_{11}e^{-M_\phi r}+p_{12}\bar{p}_{21}e^{-M_\xi r}}{3}}.
    \label{G eff}
\end{equation}
We also introduced the PPN $\gamma$, given by
\begin{equation}
    \gamma\equiv\frac{3-p_{11}\bar{p}_{11}e^{-M_\phi r}-p_{12}\bar{p}_{21}e^{-M_\xi r}}{3+p_{11}\bar{p}_{11}e^{-M_\phi r}+p_{12}\bar{p}_{21}e^{-M_\xi r}}.
    \label{PPN gamma}
\end{equation}
It is now evident from \eqref{symm far away h00}-\eqref{symm far away hij} that the parameter $\epsilon$ must satisfy
\begin{equation}\label{epsilonio}
    \abs{\epsilon} \ll \dfrac{24 G M_{\odot}}{\phi_0r^3}
\end{equation}
as long as the point $r$ lays in the region described by \eqref{metric expans}. Analogously, solar system measurements \cite{Will:2005va} constraint $\gamma\simeq 1$ and in contrast with standard metric $f(R)$ predictions \cite{Olmo:2005hc} but by close analogy with \cite{Harko:2011nh}, we see that in principle such a requirement can be fulfilled also in the presence of long range scalar interactions. Indeed, Yukawa contributes to \eqref{PPN gamma} are tuned by coefficients $p_{11}\bar{p}_{11},\,p_{12}\bar{p}_{21}$, related to the potential expansion \eqref{taylor expans W}, which can make the corrections due to the scalar field negligible also for very low masses. However, by virtue of $\text{P}\text{P}^{-1}=\text{I}$, they are not truly independent but are compelled to satisfy the condition
\begin{equation}
    p_{11}\bar{p}_{11}+p_{12}\bar{p}_{21}=1,
\end{equation}
so that we can a priori clearly distinguish two different scenarios. In the first case, taking $p_{11}\bar{p}_{11}$ and $p_{12}\bar{p}_{21}$ of the same magnitude we are forced to consider large masses for both the scalar fields in order to recover $\gamma\simeq 1$. In the second case, on the contrary, setting one of the coefficient nearly vanishing, we can retain a low mass mode which can affect astrophysical and cosmological scales. Lastly, we note that to preserve the attractive behaviour of gravity at the leading order implies, by virtue of \eqref{G eff}, that condition $\phi_0>0$ be satisfied.

\subsection{Case $W_{0,\xi}=0$}
In order to see if a configuration characterized by a low mass mode is actually attainable, we have to go back to conditions \eqref{PPNzaaro}, which besides fixing the value of  $W_{0,\phi}=\epsilon$, also implies either $\xi_0=0$ or $W_{0,\xi}=0$\footnote{We do not take into account the very special case $\xi_0=W_{0,\xi}=0$, when the matrix A has a vanishing row and the scalar fields are already decoupled, with $\delta\xi$ massless.}. Especially, when $W_{0,\xi}=0$ the matrices P, $\text{P}^{-1}$ turn out to be, respectively
\begin{equation}
    P=
    \begin{pmatrix}
    -\frac{3m_\phi^2-3m_\xi^2+U}{2\xi_0 W_{0,\phi\xi}} & 
    -\frac{3m_\phi^2-3m_\xi^2-U}{2\xi_0 W_{0,\phi\xi}} \\
    1 & 1
    \end{pmatrix}
    \label{PPN P}
\end{equation}
and
\begin{equation}
    P^{-1}=\frac{1}{2U}
    \begin{pmatrix}
    -2\xi_0 W_{0,\phi\xi} & 
    U-3m_\phi^2+3m_\xi^2 \\
    2\xi_0 W_{0,\phi\xi} & 
    U+3m_\phi^2-3m_\xi^2
    \end{pmatrix},
    \label{PPN Pinv}
\end{equation}
where $U$ is defined as
\begin{equation}
    U\equiv\sqrt{(3m_\phi^2-3m_\xi^2 )^2-4\phi_0\xi_0 W_{0,\phi\xi}^2}
    .\label{def U}
\end{equation}
Then, combining \eqref{def A} and \eqref{PPN P}-\eqref{PPN Pinv}, the masses of the decoupled scalar fields can be obtained, i.e
\begin{equation}\label{masse diagonali}
    \begin{split}
        & M^2_{\phi}\equiv\frac{1}{2}\leri{m_\phi^2+m_\xi^2 +\frac{U(\phi_0,\xi_0)}{3}} \\
        & M^2_{\xi}\equiv\frac{1}{2}\leri{m_\phi^2+m_\xi^2 -\frac{U(\phi_0,\xi_0)}{3}},
    \end{split}
\end{equation}
and the coefficients ruling the Yukawa corrections evaluated
\begin{align}
    &p_{11}\bar{p}_{11}=\frac{U+3(m_\phi^2-m_\xi^2)}{2U}\\
    &p_{12}\bar{p}_{21}=\frac{U-3(m_\phi^2-m_\xi^2)}{2U}.
    \label{p coefficients Wxi 0}
\end{align}
Now, in order to assure the existence of the function $U$ together with the positivity of \eqref{masse diagonali}\footnote{We are disregarding configurations where $M^2_\phi,\,M^2_\xi<0$, which would lead to an oscillatory behaviour for $\gamma$ (see \cite{Olmo:2005hc}) and to tachyonic instabilities in gravitational waves propagation (see Sec.~\ref{sec4}.}, the following set of inequalities must hold 
\begin{subequations}
    \begin{align}
        & \label{reality} U^2(\phi_0,\xi_0)\ge 0\\
        & \label{phiDphys} M^2_{\phi}\ge 0 \\
        & \label{xiDphys} M^2_{\xi} \ge 0,
    \end{align}
\end{subequations}
which combined yields to
\begin{equation}
    0\le U(\phi_0,\xi_0)\le 3\leri{m_\phi^2+m_\xi^2}.
    \label{correctionterm}
\end{equation}
In particular, the reality of $U$ implies
\begin{equation}
    (3m_\phi^2-3m_\xi^2 )^2-4\phi_0\xi_0 W_{0,\phi\xi}^2 \ge 0,
    \label{reality}
\end{equation}
which is always satisfied for $\xi_0<0$, while for a positive $\xi_0$ can be rewritten as 
\begin{equation}\label{xipos}
    \phi_0 W_{0,\phi\phi}+\xi_0 W_{0,\xi\xi}\ge \epsilon+2\sqrt{\phi_0\xi_0}\,|W_{0,\phi\xi}|.
\end{equation}
Instead, the inequality $U(\phi_0,\xi_0)\le 3\leri{m_\phi^2+m_\xi^2}$ states that
\begin{equation}
    \xi_0 \leri{\phi_0 \det \text{H} (W_0)-\epsilon \,W_{0,\xi\xi}}\le 0,
    \label{umax}
\end{equation}
whereas the fact that the sum of the masses of the coupled modes is compelled to be positive implies

\begin{equation}
    \phi_0 W_{0,\phi\phi}-\xi_0 W_{0,\xi\xi}\ge \epsilon.
    \label{massepos}
\end{equation}
In the case $\xi_0>0$ condition \eqref{umax} yields to
\begin{equation}\label{pipi}
    \phi_0 \det \text{H} (W_0)-\epsilon \,W_{0,\xi\xi}\le 0
\end{equation}
which, combined with \eqref{massepos}, implies the redundancy of \eqref{xipos}, hence the latter shall not be considered in the following.
Now, according the value of $U$ the mass spectrum can exhibit quite different behaviour. Indeed, when $U\neq 0$ the masses of the decoupled modes are distinct, i.e.
\begin{equation}
    M_\phi^2-M_\xi^2=\frac{U(\phi_0,\xi_0)}{3}
\end{equation}
and the spectrum is not degenerate. Then, since we are interested in peculiar scenarios where at least one scalar field is long range, we can take $U\simeq U_{max}\equiv 3(m_\phi^2+3m_\xi^2)$, for which the scalar field $\delta\xi_D$ is very tiny while the mode $\delta\phi_D$ is endowed with the mass $M_\phi^2\simeq m_\phi^2+m_\xi^2$. We also note that for $U\simeq U_{max}$, relation \eqref{umax} implies
\begin{equation}
    \phi_0 \det \text{H}(W_0)\simeq \epsilon W_{0,\xi\xi},
\end{equation}
which taken into account \eqref{ppn condition on W} leads to $W_0>0$.
\\ Now, for this arrangement of the masses the coefficients \eqref{p coefficients Wxi 0} boils down to
\begin{equation}
    p_{11}\bar{p}_{11}\simeq\frac{m_\phi^2}{m_\phi^2+m_\xi^2}\qquad p_{12}\bar{p}_{21}\simeq\frac{m_\xi^2}{m_\phi^2+m_\xi^2},
\end{equation}
and choosing $m_\phi^2\gg m_\xi^2$, i.e.
\begin{equation} \label{veryvery}
    \phi_0W_{0,\phi\phi}+\xi_0W_{0,\xi\xi}\gg \epsilon
\end{equation} we can reproduce the conditions $p_{11}\bar{p}_{11}\simeq 1$ and $p_{12}\bar{p}_{21}\simeq 0$, which are compatible with the requirement of preserve the observed dynamics at local scales, provided we properly set $m_\phi^2,\,m_\xi^2$.

\subsection{Case $\xi_0=0$}
If $\xi_0=0$ the function $f(R,\mathcal{R})$ has no linear contribution in the Palatini scalar $\mathcal{R}$ and the matrix A turns in triangular form. In this special case, the matrices P, $\text{P}^{-1}$ read as
\begin{equation}
    \text{P}=
    \begin{pmatrix}
    1 & -p \\
    0 & 1
    \end{pmatrix}
    \qquad
    \text{P}^{-1}=
    \begin{pmatrix}
    1 & p \\
    0 & 1
    \end{pmatrix},
    \label{PPN P xi0}
\end{equation}
where $p\equiv \frac{2W_{0,\xi}-\phi_0 W_{0,\phi\xi}}{\epsilon-W_{0,\xi}-\phi_0 W_{0,\phi\phi}}$.
Since now $\bar{p}_{11}=1,\;\bar{p}_{21}=0$, the matter source is not shuffled and \eqref{PPN decoupled phi}-\eqref{PPN decoupled xi} take the simpler form
\begin{align}
    &(\bigtriangleup-M_\phi^2)\delta\phi_D=-\frac{\kappa}{3}\rho
    \label{PPN decoupled phi xi0}\\
    &(\bigtriangleup-M_\xi^2)\delta\xi_D=0
    \label{PPN decoupled xi xi0},
\end{align}
with
\begin{equation}
        M_\phi^2=m_\phi^2=\frac{\phi_0 W_{0,\phi\phi}-\epsilon}{3}\qquad M_\xi^2=m_\xi^2=-\frac{W_{0,\xi}}{3}.
    \label{decoupled masses xi0}
\end{equation}
The condition for the masses of the decoupled fields to be real leads in this case to
\begin{subequations}\label{trota}
    \begin{align}
        &\phi_0 W_{0,\phi\phi}-\epsilon >0, \\
        &W_{0,\xi}<0.
    \end{align}
\end{subequations}
Solutions can be written as
\begin{align}
    &\delta\phi_D(x)=\frac{2G}{3}\int d^3x'\,\frac{\rho(x')}{\abs{x-x'}}e^{-M_\phi\abs{x-x'}}
    \label{PPN phi solution xi0}\\
    &\delta\xi_D(r)=\frac{2GM_{\odot}}{3r}e^{-M_\xi r},
    \label{PPN xi solution xi0}
\end{align}
where we normalized conveniently the expression for $\delta\xi_D$. Then, far away from the central source solutions are still given by \eqref{symm far away h00}-\eqref{symm far away hij}, where now
\begin{equation}
    G_{eff}\equiv\frac{G}{\phi_0}\leri{1+\frac{e^{-M_\phi r}-p\,e^{-M_\xi r}}{3}},
\end{equation}
and
\begin{equation}
    \gamma\equiv\frac{3-e^{-M_\phi r}+p\,e^{-M_\xi r}}{3+e^{-M_\phi r}-p\,e^{-M_\xi r}}.
    \label{PPN gamma xi0}
\end{equation}
 In this situation, since the Yukawa correction due to the scalar mode $\delta\phi_D$ cannot be tuned, likewise ordinary metric $f(R)$ case, we are compelled to consider configurations in which it is very massive and its contribute appreciable only at short scales. Conversely, properly setting the parameter $p$ we can still have a light $\delta\xi_D$ mode, provided $2W_{0,\xi}\simeq \phi_0 W_{0,\phi\xi}$, where $p\simeq 0$.

\section{Gravitational waves propagation}\label{sec4}
Now, we consider \eqref{metric expans} exact and we study the propagation of gravitational degrees of freedom on a globally flat spacetime. In this case, the consistency at the lowest order for equations \eqref{equationmetricscalartensor}, \eqref{equationphi2} and \eqref{equationxi2}, requires that the configuration $(\phi_0,\xi_0)$ be also a zero, beside a stable minimum, for the potential $W$, i.e.
\begin{equation}
    W(\phi,\xi)\simeq \frac{1}{2}\leri{W_{0,\phi\phi}\delta\phi^2+W_{0,\xi\xi}\delta\xi^2+2W_{0,\phi\xi}\delta\phi\delta\xi}.
\end{equation}
Hence, restricting our attention to the vacuum case ($T_{\mu\nu}=0$), the equation of motion for the metric field is given by 
\begin{equation}
    R_{\mu\nu}^{(1)}-\frac{1}{2}\eta_{\mu\nu}R^{(1)}=\frac{1}{\phi_0}\leri{\partial_\mu\partial_\nu-\eta_{\mu\nu}\Box}\delta\phi,
    \label{equationmetriclin}
\end{equation}
where we did not fix a priori any gauge conditions and $R_{\mu\nu}^{(1)}$ and $R^{(1)}$ are the Ricci tensor and the Ricci scalar expressed at first order in $h_{\mu\nu}$.
The linearized equations for $\phi$ and $\xi$ instead turn out to be, respectively
\begin{align}
        & \leri{\Box-\frac{\phi_0}{3}W_{0,\phi\phi}}\delta\phi-\frac{\phi_0}{3}W_{0,\phi\xi}\,\delta\xi=0\label{equation phi lin}\\
        & \leri{\Box+\frac{\xi_0}{3}W_{0,\xi\xi}}\delta\xi+\frac{\xi_0}{3}W_{0,\phi\xi}\delta\phi=0,\label{equation xi lin}
    \end{align}
where $\Box$ is the D'Alambert operator $\Box\equiv \partial_\mu\partial^\mu$.
\\ We point out that \eqref{equationmetriclin} features the same form of metric $f(R)$ theories \cite{Moretti:2019yhs}, even if the dynamical degree $\phi$ actually satisfies the remarkably different equation \eqref{equation phi lin}, which still at the linear order is coupled with the corresponding equation \eqref{equation xi lin} for $\xi$. They represent a pair of coupled wave equations for massive fields, so that extended hybrid metric-Palatini gravity seems to be characterized in vacuum by two further propagating degrees of freedom in addition to the ordinary tensorial ones. Of course, the theory could be in principle affected by instabilities concerning possible tachyonic modes, and in this respect we will see that there exists a suitable region into the parameter space of the theory, where both the modes are allowed to propagate.

\subsection{Decoupling of the wave equations}
Following the analysis made in Sec.~\ref{sec3}, the set \eqref{equation phi lin}-\eqref{equation xi lin} can be rearranged into the form
\begin{equation}
    (\text{I}_{2\times 2}\Box-\text{B})\mathbf{\Phi}=0
    \label{matrix coupled system}
\end{equation}
where now B is given by
\begin{equation}
    \text{B}\equiv\frac{1}{3}
    \begin{pmatrix}
    \phi_0 W_{0,\phi\phi} & \phi_0\,W_{0,\phi\xi} \\ \xi_0\, W_{0,\phi\xi} & -\xi_0 W_{0,\xi\xi}
    \end{pmatrix}.
    \label{def B}
\end{equation}
For $W_{0,\phi\xi} \neq 0$ the set must be turned to diagonal form, and that can be accomplished provided $\phi_0,\,\xi_0\neq 0$. Again, it is possible to rearrange \eqref{matrix coupled system} into the form
\begin{equation}
(\text{I}_{2\times 2}\Box-\text{B}_\text{D})\mathbf{\Phi_D}=0    
\end{equation}
with $\text{B}_\text{D}\equiv \text{P}^{-1}\text{A}\text{P}$, where the matrix $\text{P}$ is still given by \eqref{PPN P}, provided we replace
\begin{equation}
    m_\phi^2\equiv\frac{\phi_0}{3}W_{0,\phi\phi},\quad\quad m_\xi^2\equiv-\frac{\xi_0}{3}W_{0,\xi\xi}.
    \label{nondiag mass}
\end{equation}
Then, the following set of decoupled equations for  $\mathbf{\Phi}_D$ can be written down  
    \begin{align}
        & (\Box-M^2_{\phi})\delta\phi_D=0 \label{decoupled phi}\\
        & (\Box-M^2_{\xi})\delta\xi_D=0, \label{decoupled xi}
    \end{align}
with $M_\phi^2,\,M_\xi^2$ as in \eqref{masse diagonali}, taken into account \eqref{nondiag mass}.
Now, in order to assure that \eqref{decoupled phi}-\eqref{decoupled xi} actually describe propagating physical fields, the set of inequalities \eqref{reality}-\eqref{xiDphys} can be restated, by writing explicitly $U$ and squaring \eqref{correctionterm}, as
\begin{subequations}
    \begin{align}
        & 4\phi_0\xi_0W_{0,\phi\xi}^2 \le \leri{\phi_0 W_{0,\phi\phi}+\xi_0 W_{0,\xi\xi}}^2 \label{stability2}\\
        & 4\phi_0\xi_0 \det \text{H}(W_0)\le 0.\label{stability3}
    \end{align}
\end{subequations}
Since we are interested in stable minimum configurations \eqref{ppn condition on W}, from \eqref{stability3} it follows that $\phi_0$ and $\xi_0$ have to exhibit opposite sign. Thus, considering \eqref{massepos} for $W_{0,\phi}=0$ (i.e. $\epsilon=0$), this in turn implies that the only possible case satisfying all the criteria is 
\begin{equation}
    \phi_0>0\,\quad\quad \xi_0<0.
    \label{critical point sign}
\end{equation}
In fact, when \eqref{critical point sign} holds, relations \eqref{massepos} and \eqref{stability3} are strictly satisfied and also the squared masses of the non diagonal modes \eqref{nondiag mass} turn out to be positive. There exist, then, suitable configurations of the theory, corresponding to peculiar minima for the potential $W(\phi,\xi)$, characterized by two additional scalar degrees of freedom which propagate like linear waves on a Minkowski background. Of course, since the potential $W$ is ultimately related to the functional form $f(\cdot)$ by means of \eqref{potential V}, this selects specific classes of $f(R,\mathcal{R})$ models and the existence of one or more propagating scalar degrees could be not in general guaranteed (we remind the reader to Sec.~\ref{sec6} for details). Thus, if we restrict our attention to $f$ functions able to produce these scalar waves, we see that for $U\neq 0$ the masses corresponding to the scalar modes are distinguished, with $M_\phi^2 > M_\xi^2$ for every value of $U$ between $0$ and $U_{max}$. The specific configuration $U=U_{max}$, where the mode $\xi_D$ is predicted to become massless, has to be instead disregarded. Indeed, in that condition \eqref{stability3} would imply $\det\text{H}(W_0)=0$, where as discussed in Sec.~\ref{sec6} the scalar-tensor representation is not valid, being $\det\text{H}(f)=\infty$. We have to restrict therefore the study to the case $\det\text{H}(W_0)=\Delta U$, with $\Delta U$ a positive small parameter, where with a bit a manipulation can be shown that
\begin{equation}
  \begin{split}
   U\simeq U_{max}+2\phi_0\xi_0\Delta U,
  \end{split}
  \label{u max eps 1}
\end{equation}
 and the decoupled scalar modes are endowed with the squared masses 
 \begin{equation} \begin{split}     
 &M_\phi^2\simeq m_\phi^2+m_\xi^2+\frac{\phi_0\xi_0}{3}\Delta U \\    
 & M_\xi^2\simeq -\frac{\phi_0\xi_0}{3}\Delta U. \end{split}
 \label{u max eps 2}
 \end{equation}
\\When $U=0$, instead, by virtue of \eqref{def U} and \eqref{critical point sign}, the following constraints have to be separately satisfied:
\begin{equation}
    W_{0,\phi\xi}=0,\quad\quad \phi_0 W_{0,\phi\phi}+\xi_0 W_{0,\xi\xi}=0,
    \label{cond U nullo}
\end{equation}
and it follows from \eqref{def A} that we actually deal with a system already decoupled. The procedure involving the definition of $U$, therefore, is not well grounded, and we cannot simply perform the limit of $U \rightarrow 0$ in \eqref{masse diagonali}, that would result in the degenerate spectra
\begin{equation}
    M_\phi^2=M_\xi^2=m^2\equiv\phi_0 W_{0,\phi\phi}=-\xi_0 W_{0,\xi\xi}.
\end{equation}
Rather, if $W_{0,\phi\xi}=0$, we just retain \eqref{nondiag mass}, where the masses could be in principle different: The mass spectrum is affected by a discontinuity for $U=0$, where the masses of the actual physical modes do not coincide with the values predicted by \eqref{masse diagonali}.

\section{Geodesic deviation}\label{sec5}
\begin{figure}
\begin{center}
\includegraphics[width=1\columnwidth]{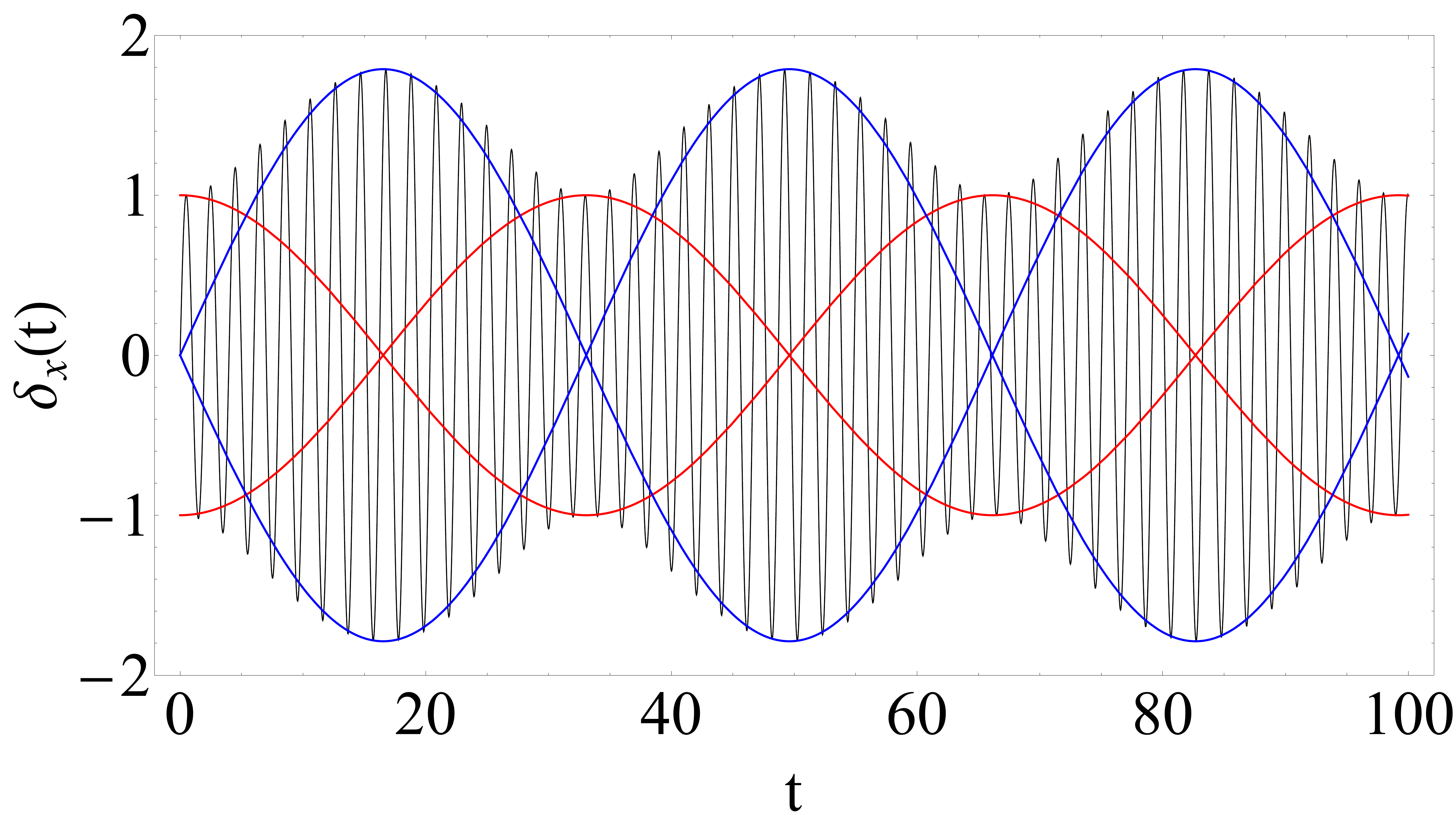}
\caption{The perturbation $\delta_x$ in function of the time $t$ due to superposition of the scalar breathing modes when $U\simeq 0$. The carrying signals with beating frequency $\Delta\Omega$ are also shown (color online). We set $\Delta\Omega/\Omega\simeq 0.03$.}
\label{Beating}
\end{center}
\end{figure}

In order to analyze the phenomenology of gravitational waves in extended hybrid theories, we can evaluate, via the geodesic deviation equation, the perturbations induced by the scalar modes on a sphere of test masses. These are displayed along with the tensorial degrees in
\begin{equation}
    \frac{\partial^2 X_i}{\partial t^2}=-R^{(1)}_{i0j0}X^j,
    \label{geodesic eq base}
\end{equation}
where we introduce the vector \begin{equation}
    \vec{X}=(x_0+\delta_x,y_0+\delta_y,z_0+\delta_z),
    \label{vec}
\end{equation}
denoting the separation between two nearby geodesics, with $x_0$ and $\delta_x$ indicating the rest position and the displacement of order $\mathcal{O}(h)$ induced by waves, respectively\footnote{Analogously for $y,z$.}.
Then, following \cite{Flanagan:2005yc}, we introduce for the metric perturbation $h_{\mu\nu}$ the generic decomposition for a symmetric tensor of rank two, i.e.
\begin{equation}
\begin{split}
h_{00}&=2\alpha \\
h_{0i}&=\beta_i+\partial_i \chi \\
h_{ij}&=h^{TT}_{ij} + \dfrac{1}{3}H\delta_{ij}+\partial_{(i}\epsilon_{j)}+\left(\partial_i\partial_j-\dfrac{1}{3}\delta_{ij}\bigtriangleup\right)\lambda,
\end{split}
\label{decomp h}
\end{equation}
with $\delta_{ij}$ the Kronecker delta, $\bigtriangleup=\partial_i\partial^i$ the Laplacian operator and symmetrization given by $A_{(ij)}\equiv\frac{1}{2}(A_{ij}+A_{ji})$.
The irreducible parts introduced in \eqref{decomp h} are accompanied by the conditions
\begin{equation}
\begin{split}
\partial^i \beta_i&=0 \\
\partial^i h^{TT}_{ij}&=0 \\
\eta^{ij}h^{TT}_{ij}&=0\\
\partial^i \epsilon_i &=0,
\end{split}
\end{equation}
which, as stressed in \cite{Flanagan:2005yc} (see also \cite{Weinberg:2008zzc} for the curved background case), are required in order to preserve the uniqueness and the consistency of the procedure. By means of these quantities we can then introduce the set of variables
\begin{equation}
\begin{split}
\Pi &= -\alpha + \partial_t{\chi}-\dfrac{1}{2}\partial_t^2{\lambda} \\
\Theta &= \dfrac{1}{3} \left ( H-\bigtriangleup \lambda\right)\\
\Xi_i &= \beta_i-\dfrac{1}{2}\partial_t{\epsilon}_i,
\end{split}
\end{equation}
which turns out to be invariant, together with $h_{ij}^{TT}$, under a linear gauge transformation.
\\ Now, as it was outlined in \cite{Moretti:2019yhs} for metric $f(R)$ theories, it is possible to rearrange the linearized equation for the metric \eqref{equationmetriclin} into the form
\begin{subequations}
	\begin{align}
	\bigtriangleup \Pi_\phi&=0\\
	\bigtriangleup \Theta_\phi &=0 \\
	\bigtriangleup \Xi_i &=0\\
	\Box h^{TT}_{ij}&=0,
	\end{align}
	    \label{equation metric gauge inv}
\end{subequations}
where we introduced the modified static degrees
\begin{equation}
    \Pi_\phi\equiv\Pi+\frac{1}{2}\frac{\delta\phi}{\phi_0},\quad\quad \Theta_\phi\equiv\Theta+\frac{\delta\phi}{\phi_0}.
    \label{def new static}
\end{equation}
From \eqref{equation metric gauge inv}, it is evident that beyond the scalar degrees discussed in Sec.~\ref{sec4}, we retain the standard tensorial modes for the metric $h_{\mu\nu}$. Moreover, it is worth noting that with respect to the discussion in \cite{Moretti:2019yhs}, the scalar field involved into the definition \eqref{def new static} does not represent a proper degree of freedom. Indeed, the quantity $\delta\phi$ is actually related by means of \eqref{PPN P} to the diagonal scalar modes $\leri{\delta\phi_D,\,\delta\xi_D}$, i.e.
\begin{equation}
\begin{split}
    \delta\phi=&-\frac{3m_\phi^2-3m_\xi^2+U}{2\xi_0 W_{0,\phi\xi}}\;\delta\phi_D\\
    &-\frac{3m_\phi^2-3m_\xi^2-U}{2\xi_0 W_{0,\phi\xi}}\;\delta\xi_D.
\end{split}
\label{delta phi in diag modes}
\end{equation}
Therefore, when we look at the  components of the linearized Riemann tensor entering \eqref{geodesic eq base}, these can be expressed in gauge invariant variables as
\begin{equation}
    R^{(1)}_{i0j0}=-\dfrac{1}{2}\partial_t^2{h}_{ij}^{TT}+\partial_{(i}\partial_t{\Xi}_{j)}+\partial_i\partial_j\Pi-\dfrac{1}{2}\delta_{ij}\partial_t^2{\Theta},
\end{equation}
which can be rewritten, neglecting the static contributions and taking into account \eqref{def new static} and \eqref{delta phi in diag modes}, like:
\begin{equation}
    \begin{split}
        R^{(1)}_{i0j0}=-\dfrac{1}{2}\partial_t^2{h}_{ij}^{TT}&+\mathcal{U}^{(+)}\leri{\partial_i\partial_j-\delta_{ij}\partial_t^2}\delta\phi_D\\
        &+\mathcal{U}^{(-)}\leri{\partial_i\partial_j-\delta_{ij}\partial_t^2}\delta\xi_D,
    \end{split}
    \label{Riemann components}
\end{equation}
with $\mathcal{U}^{(+)},\,\mathcal{U}^{(-)}$ given by, respectively
\begin{equation}
    \mathcal{U}^{(\pm)}\equiv\frac{3m_\phi^2-3m_\xi^2\pm U}{4\phi_0\xi_0 W_{0,\phi\xi}}.
    \label{def u corsivo}
\end{equation}
Now, leaving aside the tensorial degrees contained in $h_{ij}^{TT}$ and choosing the $z$ axis coincident with the direction of propagation of the waves, the scalar degrees can be described by
\begin{equation}
\begin{split}
    \delta\phi_D &= A_\phi \sin\leri{{\Omega_\phi t -K_\phi z}}\\ \delta\xi_D &=A_\xi \sin{\leri{\Omega_\xi t -K_\xi z}},
\end{split}
    \label{phi xi wave sol}
\end{equation}
with frequencies
\begin{equation}
    \Omega_\phi=\sqrt{K^2_\phi+M^2_\phi} \qquad \Omega_\xi=\sqrt{K_\xi^2+M^2_\xi},
    \label{phi xi freq}
\end{equation}
wave vectors fixed in
\begin{equation}
    K^\mu_{\phi,\xi}=\left(\sqrt{K^2_{\phi,\xi}+M^2_{\phi,\xi}},0,0,K_{\phi,\xi}\right),
\end{equation}
and $A_\phi,\,A_\xi$ the amplitudes of the waves. 
Thus, the geodesic deviation equation takes the form:
\begin{equation}
\begin{split}
    &\partial_t^2{\delta}_x\simeq  -x_0 \leri{ \mathcal{U}^{(+)}\leri{K^2_\phi+M^2_\phi} \delta\phi_D+\mathcal{U}^{(-)}\leri{K^2_\xi+M^2_\xi}\delta\xi_D}\\
     &\partial_t^2{\delta}_y\simeq -y_0 \leri{ \mathcal{U}^{(+)}\leri{K^2_\phi+M^2_\phi} \delta\phi_D+\mathcal{U}^{(-)}\leri{K^2_\xi+M^2_\xi}\delta\xi_D}\\
      &\partial_t^2{\delta}_z\simeq -z_0 \leri{ \mathcal{U}^{(+)}M^2_\phi \delta\phi_D+\mathcal{U}^{(-)}M^2_\xi\delta\xi_D},
   \end{split}    
   \label{geodesic}
\end{equation}
where we disregarded terms of order $\mathcal{O}(h^2)$.
\\ By close analogy with the discussion in \cite{Moretti:2019yhs,Liang:2017ahj}, we see that both the scalar degrees are able to induce two type of polarizations. In fact, they are separately responsible for a breathing mode on the transverse plane $xy$, as well as for a longitudinal excitation along the direction of propagation of the wave. Moreover, the corresponding polarizations are modulated for $U\neq 0$  by factors of distinct magnitude, and $\delta\phi_D,\,\delta\xi_D$ propagate with different speed. In particular, when $U\rightarrow U_{max}$, by virtue of \eqref{u max eps 1} and \eqref{u max eps 2} the longitudinal contribute of $\delta\xi_D$ turns out to be or order $\varepsilon$, i.e.
\begin{equation}
\begin{split}
    & \mathcal{U}^{+}M_\phi^2\simeq \sgn (W_{0,\phi\xi}) \frac{m_\phi^2+m_\xi^2}{2\xi_0}\sqrt{\frac{W_{0,\phi\phi}}{W_{0,\xi\xi}}}+\mathcal{O}(\varepsilon) \\
    & \mathcal{U}^{-}M_\xi^2\simeq -\sgn (W_{0,\phi\xi}) \frac{\xi_0}{6}\sqrt{\frac{W_{0,\xi\xi}}{W_{0,\phi\phi}}}\;\varepsilon.
\end{split}
\end{equation}
In this case, therefore, it mostly affects the geodesic deviation as a breathing on the plane $xy$ and the longitudinal polarization is almost entirely due to the massive mode $\delta\phi_D$.
\\ Conversely, when $U\rightarrow 0$ the angular frequencies of the scalar modes are very close to each other, and we expect that typical interference patterns between waves, i.e. beatings, could take place. That can be considered a very distinctive marker of gravitational wave propagation in generalized hybrid metric-Palatini theories, absent in ordinary metric $f(R)$ gravity, with specific phenomenological implications.
\\ Thus, let us write for $U\simeq 0$ the solution of \eqref{geodesic} as
\begin{equation}
\begin{split}
    &\delta_x(t)\simeq A_\phi^B\sin\leri{\Omega_\phi t}+A_\xi^B\sin\leri{\Omega_\xi t}\\
    &\delta_y(t)\simeq A_\phi^B\sin\leri{\Omega_\phi t}+A_\xi^B\sin\leri{\Omega_\xi t}\\
    &\delta_z(t)\simeq A_\phi^L\sin\leri{\Omega_\phi t}+A_\xi^L\sin\leri{\Omega_\xi t},
   \end{split} 
    \label{geodesic eff amplit}
\end{equation}
where we set $z=0$ and effective amplitudes $A_{\phi,\xi}^{B,L}$ given by
\begin{equation}
    \begin{split}
        &A_\phi^B\equiv\mathcal{U}^{(+)}A\qquad A_\phi^L\equiv\frac{\mathcal{U}^{(+)}M^2_\phi}{k^2+M_\phi^2}A\\
        &A_\xi^B\equiv\mathcal{U}^{(-)}A\qquad A_\xi^L\equiv\frac{\mathcal{U}^{(-)}M^2_\xi}{k^2+M_\xi^2}A,
    \end{split}
    \label{def amplit eff}
\end{equation}
with $A_\phi\sim A_\xi =A$ and $K_\phi\sim K_\xi = k $. After a bit of manipulation, it is possible to recast\footnote{We just report the result for $\delta_x$. Similar considerations hold for $\delta_y,\,\delta_z$.} \eqref{geodesic eff amplit} like
\begin{equation}
    \begin{split}
        \delta_x\simeq &\leri{A_\phi^B+A_\xi^B}\cos\leri{\Delta\Omega t}\sin\leri{\Bar{\Omega}t} \\
        &+\leri{A_\phi^B-A_\xi^B}\sin\leri{\Delta\Omega t}\cos\leri{\Bar{\Omega}t},
    \end{split}
    \label{prostaferesi}
\end{equation}
where we defined $\Delta\Omega\equiv\frac{ \Omega_\phi-\Omega_\xi}{2}$ and $\bar{\Omega}\equiv\frac{\Omega_\phi+\Omega_\xi}{2}$. Then, the perturbation described by \eqref{prostaferesi} represents a superposition of two waves of frequencies $\bar{\Omega}$ with a phase shift of $\pi/2$, both modulated by the beating frequency $\Delta\Omega$ (Fig.~\ref{Beating}), and relative amplitude $A_\phi^B-A_\xi^B>A_\phi^B+A_\xi^B$. 
\\ Finally, when $W_{0,\phi\xi}=0$ the set of equations \eqref{equation phi lin}-\eqref{equation xi lin} is naturally decoupled, and the transformation \eqref{delta phi in diag modes} is no longer necessary. In this case the relevant components of the Riemann are given by
\begin{equation}
        R^{(1)}_{i0j0}=-\dfrac{1}{2}\partial_t^2{h}_{ij}^{TT}-\dfrac{1}{2\phi_0}\leri{\partial_i\partial_j - \delta_{ij} \partial_t^2}\delta \phi,
\end{equation}
and we see that only $\delta \phi$ enters the geodesic deviation. Therefore, the phenomenology described is identical to that descending from the scalar-tensor formulation of metric $f(R)$ theories, i.e.
\begin{equation}
\begin{split}
    &\partial_t^2{\delta}_x\simeq \dfrac{x_0}{2\phi_0} \leri{k_\phi^2+m^2_\phi}\,\delta \phi \\
     &\partial_t^2{\delta}_y\simeq\dfrac{y_0}{2\phi_0} \leri{k_\phi^2+m^2_\phi}\,\delta \phi \\
      &\partial_t^2{\delta}_z\simeq \dfrac{z_0}{2\phi_0} m^2_\phi\,\delta \phi .
   \end{split}    
   \label{geodesic W0}
\end{equation}
Nevertheless, even if $\delta\xi$ does not appear explicitly in \eqref{geodesic W0}, we cannot infer that the functional dependence of $f(\cdot)$ on $\mathcal{R}$ have no phenomenological implications. Indeed, since $\phi$ is actually the combination of $\psi$ and $\xi$, we clearly see that both contributions of $f$ from $R$ and $\mathcal{R}$ concur in determining the effects of \eqref{geodesic W0}.

\section{Constraints on the form of $f(R,\mathcal{R})$}\label{sec6}
In this section we analyze in detail the implications onto the form of the function $f(R,\mathcal{R})$ of conditions discussed in Sec.~\ref{sec4}-\ref{sec5}. In particular, we are interested in establishing clear relations between derivatives of the potential $W$ and corresponding derivatives of the function $f$ with respect to the curvatures $R$ and $\mathcal{R}$. Then, in order to do that, it is useful to express the derivatives of $W(\phi,\xi)$ in terms of derivatives of $V(\psi,\xi)$, i.e.
\begin{align}
    & W_\phi=V_\psi \frac{\partial \psi}{\partial \phi}=V_\psi \\
    & W_\xi = V_\psi \frac{\partial \psi}{\partial \xi}+V_\xi=-V_\psi+V_\xi,
\end{align}
when we considered $W(\phi,\xi)=V(\psi(\phi,\xi),\xi)$. It follows that second order derivatives are given by
\begin{align}
    & W_{\phi\phi}=V_{\psi\psi}\\
    & W_{\xi\xi}=V_{\psi\psi}-2V_{\psi\xi}+V_{\xi\xi}\\
    & W_{\phi\xi}=V_{\psi\xi}-V_{\psi\psi},
\end{align}
which can be further rewritten, taking into account definitions of the potential $V$, as
\begin{equation}
\begin{split}
    & W_{\phi\phi}=R_{\psi}\\
    & W_{\xi\xi}=R_{\psi}-2R_{\xi}+\mathcal{R}_{\xi}\\
    & W_{\phi\xi}=R_{\xi}-R_{\psi}.
    \label{from W to R}
\end{split}
    \end{equation}
Now, since $\psi$ and $\xi$ are function of $R$ and $\mathcal{R}$ by means of the first derivatives of $f$, in evaluating \eqref{from W to R} we can apply the inverse function theorem for the two dimensional case, leading to
\begin{equation}
    \begin{split}
    &V_{\psi\psi}=R_\psi=\frac{f_{\mathcal{R}\mathcal{R}}}{\det\text{H}(f)}\\
    &V_{\psi\xi}=R_\xi=\mathcal{R}_\psi=-\frac{f_{R\mathcal{R}}}{\det\text{H}(f)}\\
    &V_{\xi\xi}=\mathcal{R}_\xi=\frac{f_{RR}}{\det\text{H}(f)},    
    \end{split}
    \label{from V to f}
\end{equation}
where we use the fact that the Jacobian matrix of the transformation relating $(\psi,\,\xi)$ to $(R,\,\mathcal{R})$ coincides with the Hessian matrix for $f$.
We can then express the determinant of the Hessian matrix of $W$ in terms of derivatives of $f$, that is
\begin{equation}
    \det \text{H}(W)=\det \text{H}(V)=\frac{1}{\det \text{H}(f)}.
    \label{det W to det f}
\end{equation}
In the continuing we will evaluate these quantities at background values $\phi_0,\xi_0$ and we see that condition $\det \text{H}(f)\neq 0$, required for scalar-tensor representation to exist, guarantees that $(R,\mathcal{R})$ could be solved for $(\phi,\xi)$ and therefore computed in $(\phi_0,\xi_0)$. Furthermore, from \eqref{det W to det f} it is evident that we have to disregard from the analysis configurations with $\det \text{H}(W_0)= 0$, corresponding to $\det \text{H}(f)=\infty$, where a propagating massless mode is theoretically predicted for gravitational waves (see Sec.~\ref{sec5}).

\subsection{Post parameterized Newtonian corrections}
In studying the PPN corrections we made the following general assumptions
\begin{equation}
    \begin{cases}
    & \det \text{H}(W_0)>0 \\
    & W_{0,\phi\phi}>0 \\
    & W_{0,\xi\xi}>0 \\
    & \phi_0 >0,
    \end{cases}
    \label{ppn conditions on f}
\end{equation}
which can be translated by means of \eqref{from W to R} and \eqref{from V to f} in conditions on the derivatives of the function $f$. This results in
\begin{equation}
    \begin{cases}
    & \det \text {H}(f_0)>0 \\
    &  f_{0,\mathcal{R}\mathcal{R}}>0\\
    & f_{0,RR}+f_{0,\mathcal{R}\mathcal{R}}+2f_{0,R\mathcal{R}}>0 \\
    & f_{0,R}+f_{0,\mathcal{R}}>0.
    \end{cases}
\end{equation}
where we used the definitions of $\phi$ and introduced, by analogy with $W$, the subscript 0 also for $f$. We see that second order derivatives constitute an independent subsystem of inequalities, whose solution is given by
\begin{equation}
    f_{0,RR}>0,\;\; f_{0,\mathcal{R}\mathcal{R}} > 0,\;\; \abs{f_{0,R\mathcal{R}}} < \sqrt{f_{0,RR}f_{0,\mathcal{R}\mathcal{R}}},
\end{equation}
while first order derivatives do not require further manipulations. In the continuing, we will investigate in detail other conditions required in each case discussed in Sec.~\ref{sec3}.
\subsubsection{Case $W_{0,\xi}=0$, $\xi_0<0$}
The condition of reality for the function $U$ \eqref{reality} always holds, while inequalities \eqref{umax} and \eqref{massepos} can be reformulated as
\begin{subequations}
\begin{align}
    & \dfrac{f_{0,R}-|f_{0,\mathcal{R}}|}{f_{0,RR}+f_{0,\mathcal{R}\mathcal{R}}+2f_{0,R\mathcal{R}}}\ge \epsilon\label{gigi} \\
    & \dfrac{f_{0,R}f_{0,\mathcal{R}\mathcal{R}}+|f_{0,\mathcal{R}}|\leri{f_{0,RR}+2f_{0,R\mathcal{R}}}}{\det \text{H}(f_0)}\ge \epsilon,
\end{align}
\end{subequations}
where we have outlined the negative sign of $f_{0,\mathcal{R}}=\xi_0$.
 The requirement $m^2_\phi \gg m^2_\xi$ that we impose in order to correctly reproduce the Newtonian limit can be restated as 
\begin{equation}\label{moltomaggiore}
   \dfrac{f_{0,R}f_{0,\mathcal{R}\mathcal{R}}-|f_{0,\mathcal{R}}|\leri{f_{0,RR}+2f_{0,\mathcal{R}\mathcal{R}}+2f_{0,R\mathcal{R}}}}{\det \text{H}(f_0)}\gg \epsilon.
\end{equation}

\subsubsection{Case $W_{0,\xi}=0$, $\xi_0>0$}
Conditions \eqref{massepos}, \eqref{pipi} and \eqref{veryvery} can be reformulated as follows
\begin{equation}
\begin{cases}
\epsilon W_{0,\phi\phi}\ge \xi_0\det \text{H}(W_0) \\
\epsilon W_{0,\xi\xi}\ge \phi_0\det \text{H}(W_0) \\
\phi_0  W_{0,\phi\phi} +\xi_0 W_{0,\xi\xi}\gg \epsilon,
\end{cases}
\end{equation}
where in the first inequality we have omitted $\epsilon^2$ terms.
It is immediate to translate this set of inequalities in terms of constraints on the derivatives of the function $f$, yielding to
\begin{equation}
    \begin{cases}
    0< \dfrac{f_{0,\mathcal{R}}}{f_{0,\mathcal{R}\mathcal{R}}}\le \epsilon \\
  0<\dfrac{f_{0,R}+f_{0,\mathcal{R}}}{f_{0,RR}+f_{0,\mathcal{R}\mathcal{R}}+2f_{0,R\mathcal{R}}}\le\epsilon \\
  \dfrac{f_{0,R}f_{0,\mathcal{R}\mathcal{R}}+f_{0,\mathcal{R}}\leri{f_{0,RR}+2f_{0,\mathcal{R}\mathcal{R}}+2f_{0,R\mathcal{R}}}}{\det \text{H}(f_0)}\gg \epsilon.
    \end{cases}
\end{equation}
\subsubsection{Case $\xi_0=0$}
Conditions \eqref{trota} reads, in the special setting $p\simeq 0$, as 
\begin{equation}
    \begin{cases}
    \dfrac{f_{0,\mathcal{R}\mathcal{R}}(f_{0,R}+f_{0,\mathcal{R}})}{\det \text{H}(f_0)}>\epsilon \\
    f_{0,\mathcal{R}\mathcal{R}}+f_{0,R\mathcal{R}}>0.
    \end{cases}
\end{equation}
 In order to reproduce PPN corrections which are compatible with local measurements we have to impose $M_\phi \bar{r} \gg 1$, with $\bar{r}$ much smaller than the size of the source. By combining this request with \eqref{epsilonio} we get 
\begin{equation}
    \dfrac{f_{0,\mathcal{R}\mathcal{R}}(f_{0,R}+f_{0,\mathcal{R}})}{\det \text{H}(f_0)}\gg \epsilon + 3 \leri{\dfrac{\phi_0\epsilon}{24GM_\odot}}^{\frac{2}{3}}.
\end{equation}
\subsection{Gravitational wave modes}
As we saw in Sec.~\ref{sec5}, for stable minima of $W(\phi,\xi)$ two additional massive scalar modes are expected to propagate, provided the set \eqref{ppn conditions on f} of inequalities hold, along with the additional condition $\xi_0<0$.
In terms of $f$ this leads to
\begin{equation}
    \begin{cases}
    & \det \text {H}(f_0)>0 \\
    &  f_{0,\mathcal{R}\mathcal{R}}>0\\
    & f_{0,RR}+f_{0,\mathcal{R}\mathcal{R}}+2f_{0,R\mathcal{R}}>0 \\
    & f_{0,R}+f_{0,\mathcal{R}}>0.\\
    & f_{0,\mathcal{R}}<0,
    \end{cases}
    \label{constraints on f GW}
\end{equation}
 and as in the PPN case the constraints on the second derivatives are satisfied if 
\begin{equation}
    f_{0,RR}>0,\;\; f_{0,\mathcal{R}\mathcal{R}} > 0,\;\; \abs{f_{0,R\mathcal{R}}} <  \sqrt{f_{0,RR}f_{0,\mathcal{R}\mathcal{R}}}.
\end{equation}
Finally, for the special setting $W_{0,\phi\xi}=0$, where the scalar modes are already decoupled and equipped with massed as in \eqref{nondiag mass}, conditions \eqref{constraints on f GW} are endowed with the further requirement $f_{0,R\mathcal{R}}+f_{0,\mathcal{R}\mathcal{R}}=0$. That, plugged back into \eqref{constraints on f GW}, leads then to $f_{0,RR}>f_{0,\mathcal{R}\mathcal{R}}=-f_{0,R\mathcal{R}}>0$.

\section{Concluding remarks}\label{sec7}
Extended hybrid metric-Palatini theories are promising generalizations of the two main approaches to the study of $f(R)$ gravity, and the most intriguing feature of these models is certainly the presence of two dynamical scalar fields non minimally coupled to gravity. In fact, this enrichment of the dynamical structure can be used, in principle, to remove technical and conceptual problems, common to both the metric and the Palatini approach, that arise when one tries to mimic dark matter or dark energy effects without spoiling Solar System tests. In this work we investigated the weak field limit of the theory in its scalar-tensor formulation, by analyzing the first PPN order and the gravitational wave propagation. In both cases we found the two scalar fields solve coupled equations, and the decoupled scalar fields are shown to be massive, with masses that vary in a range determined by the potential $W$. Particularly, the mass spectrum spans continuously the interval that goes from two modes having nearly the same mass to the case in which one field has the maximum mass (again determined by $W$) while the other field is massless. In this respect, we clarified how this peculiar setting depends crucially on the parameter $\epsilon$ quantifying the departure at the background level from Minkowski spacetime. We showed, in fact, that when the propagation of gravitational scalar waves is addressed, corresponding to $\epsilon=0$, the configuration where $\delta\xi_D$ is massless is not actually feasible, in that scalar tensor representation is not attainable. With regard to the PPN expansion we found that the presence of the massive fields implies that the parameters $G_{eff}$ and $\gamma$ acquire Yukawa-like corrections. The intensity of these modifications is governed by coefficients that can be tuned through specific constraints on the potential function $W$. We claim that it is possible to make corrections in the expressions of the PPN parameters small enough to stay within the constraints of current Solar System tests, still having the presence of a scalar massive field light enough to act as dark matter on galactic scales. This can be accomplished by choosing $W$ such that the masses of the scalar fields are widely separated: with this expedient the mass of the heavier can be set to a value that implies the suppression of the relative exponential factor over a convenient scale, while the lighter can be forced to have a decay length comparable with galactic scales. The correction relative to the light scalar in the expressions of $G_{eff}$ and $\gamma$ can be made small enough through a precise choice on the corresponding coefficient. For what concerns the gravitational wave study, we performed a linear metric approximation around a Minkowski background, hence we restricted the dynamics of the scalar fields to small oscillations around a local minimum of $W$. We showed that the decoupled fields solve two independent Klein-Gordon equations with masses varying in the above mentioned range. The analysis of the phenomenology associated to the scalar fields, performed via the geodesic deviation equation for a sphere of test particles, demonstrated that each field is detectable as the superposition of two independent polarizations, namely a breathing plus a longitudinal mode. It must be stressed that such a finding cannot be claimed to be a real marker for this specific model: indeed in \cite{Liang:2017ahj,Moretti:2019yhs} is shown that in the metric formalism the only additional scalar field is responsible for the same mixture of polarizations, whereas in \cite{Montani:2018iqd} is demonstrated that in General Relativity gravitational waves travelling in molecular media, like  galaxies, are expected to contain additional mode characterized by the same feature. The striking peculiarity of this model is instead the fact that the two scalar fields can mutually interact and produce beatings. We studied this phenomenon in the special case of nearly degenerate masses, in which the beating frequency is much smaller than the signal frequency. However, the same feature should be detectable for any value of the masses in the allowed range, at least until the lighter scalar can be properly distinguished from the massless tensorial degrees, in which case we expect their mutual interaction as in \cite{Bombacigno:2018tih}. Finally, we established precise relations between the constraints on the potential $W(\phi,\xi)$, obtained in analyzing PPN and gravitational wave settings, and the form of the function $f(R,\mathcal{R})$. These resulted in a set on inequalities connecting second and first derivatives of $f$ with respect both the curvatures. In particular, they could be used in principle for the construction of a definite model meeting the requirements that are necessary to mimic dark matter effects and pass the Solar System tests, paying special attention to already known potentials (as discussed in \cite{Rosa:2017jld}) characterized by accelerating cosmological solutions as well.

\end{document}